\documentstyle[aps,epsfig]{revtex}

\newcommand{\ffig}[4]{\begin{figure}[hp]\vfill\begin{center}
\mbox{\epsfig{figure=#1,height=#2}}\caption{#3}\label{#4}
\end{center}\vfill\end{figure}}

\newcommand{\PO}{\rm l \! P }
\newcommand{\RO}{\rm l \! R }
\newcommand{\xpom}{x_{\PO} }

\newcommand{\GeV}{\mbox{\rm ~GeV~}}

\begin{document}

\title{QCD analysis of the diffractive structure function $F_2^{D(3)}$}
\author{C. Royon\thanks{%
 Service de Physique des Particules,
CE-Saclay, F-91191 Gif-sur-Yvette Cedex, France;
Brookhaven National Laboratory, Upton, New York, 11973;
University of Texas, Arlington, Texas, 76019},
L. Schoeffel \thanks{%
 Service de Physique des Particules,
CE-Saclay, F-91191 Gif-sur-Yvette Cedex, France},
J.Bartels\thanks{II. Institut f\"ur Theoretische Physik, Universit\"at 
Hamburg, Luruper Chaussee 149, D-22761 Hamburg},
 H.Jung\thanks{%
University of Lund, Soelvegatan 14, 223 62 Lund, Sweden},
R.Peschanski \thanks{%
 Service de Physique Th\'eorique,
CE-Saclay, F-91191 Gif-sur-Yvette Cedex, France}
}
\maketitle

\begin{abstract}
The proton diffractive structure function $F_2^{D(3)}$ measured in the H1 and 
ZEUS experiments at HERA is analyzed in terms of both Regge phenomenology and 
perturbative QCD evolution. A new  method  determines the values of the  Regge 
intercepts in ``hard'' diffraction, confirming a higher value of the Pomeron 
intercept than for soft physics. The data are well described by a QCD analysis
in which point-like parton distributions, evolving according to the DGLAP 
equations, are assigned to the leading and sub-leading Regge exchanges.
The gluon distributions are found to be quite different for H1 and ZEUS. A {\it 
global fit} analysis, where a higher twist component is taken from models, allows
us
to use data in the whole available range in diffractive mass and gives a stable answer for 
the 
leading twist contribution. We give sets of quark and gluon parton distributions for the 
Pomeron, and  predictions for the charm and the longitudinal  proton diffractive 
structure function from the QCD fit. An extrapolation to the Tevatron range is 
compared with CDF data on single diffraction. Conclusions on factorization 
breaking depend critically whether H1 (strong violation) or ZEUS (compatibility 
at low $\beta$) fits are taken into account. 
\end{abstract}

\section{Introduction}
It is now experimentally well established  at HERA \cite{f2d94,zeus} 
that a substantial fraction of $ep$ events is contributable to diffraction,
i.e. color singlet exchange, 
initiated by a highly virtual photon. Starting with the pioneering theoretical 
work of Ref.\cite{ingelman}, the idea of a point-like structure of the Pomeron 
exchange opens the way to the determination of its parton (quark and 
gluon) distributions, where the Pomeron point-like structure 
can be treated in a similar way as (and compared to) the proton one. Indeed, 
leading twist contributions to the proton diffractive structure functions can be 
defined by factorization properties \cite{soper} in much the same way as for 
the full proton structure functions themselves. As such, they should obey DGLAP 
evolution equations \cite{dglap}, and thus allow for perturbative predictions of 
their 
$Q^2$ evolution.

Indeed, such parton distributions are very useful to investigate the 
difficult and longstanding
problem of the nature of the Pomeron Regge singularity. On a phenomenological 
ground, 
they are the basis of MC simulations like RAPGAP \cite {jung} and they 
give a comparison  basis with ``hard'' diffraction at Tevatron, where the 
 factorization properties are not expected to be valid.
Moreover, the study of diffractive parton distributions is also a challenge for 
the discussion of different approaches and models where other than leading twist 
contributions can be present in hard diffraction. 
Indeed, there exists strong presumptive evidence that higher twist effects may 
be quite important in diffractive processes contrary to non-diffractive 
ones which do not  require (at least at not too small $Q^2$) such 
contributions.  In fact, there are  
models which incorporate non-negligible 
contributions from    
higher twist components, especially for relatively small masses  of the 
diffractive 
system. One of our goals  is to 
take into account this peculiarity of diffractive processes. 

The purpose of this paper is to derive parton distributions
of the Pomeron from QCD fits of diffractive DIS cross-sections
determined at HERA \cite{f2d94,zeus}. By comparison with previous determinations 
contained in the experimental papers, we introduce some new techniques of 
parametrizations and discuss various points which have appeared in the
domain:

-the precise determination of the effective Pomeron and Reggeon 
trajectories  in Regge fits
  
-the determination of the  parton distributions at Next-to-Leading Order (NLO) in 
the Pomeron comparing two methods, either using cuts  in order to damp the 
expected higher twist contributions
or by a {\it global fit}  including a  higher twist contribution taken from 
models

-the differences between parton distributions derived from ZEUS and H1 data

-the need to evaluate precisely   
the gluon content of the Pomeron and its error
since it has strong impact on the discussion 
of diffractive charm and longitudinal diffractive structure functions

-the extrapolation of the diffractive structure function to
single diffraction in $p \bar{p}$ collisions at Tevatron
could give a precise evaluation of the  factorization breaking.

Our paper is structured in the following way: In section {\bf II}, we show Regge 
fits 
and the 
corresponding determination of intercepts from H1 ({\bf II-A}) and ZEUS ({\bf 
II-B}) data using a new method described in the appendix {\bf A1}. In section 
{\bf 
III} we use the same data but with  additional cuts including 
a high $\beta$ cut to 
determine the parton distributions of the Pomeron in a NLO DGLAP evolution 
framework. The possibility of two different fits for H1 ({\bf III-A}) and a 
remaining 
difference between H1 and ZEUS ({\bf III-B}) are thoroughly examined. In section 
{\bf IV}, we 
perform a new {\it global fit} analysis by removing the high $\beta$ cut, 
and using models to provide 
the higher twist contribution. In {\bf 
IV-A} we check the stability of the determination of the leading twist parton 
distributions of the Pomeron. In {\bf IV-B} we discuss the obtained gluon 
density in the Pomeron (for H1 and ZEUS fits) and examine the corresponding 
predictions for the 
diffractive charm ({\bf IV-C}) and longitudinal ({\bf IV-D}) components. The 
final selection of parameters for H1 and ZEUS fits is given in appendix {\bf 
A2}. In  section {\bf V} we discuss the important consequence of these 
determinations for single diffraction  at the Tevatron, as measured by the 
CDF collaboration \cite{cdf}. We see that the conclusions on factorization 
breaking are markedly different for the H1 and ZEUS determinations. 
Conclusions  and outlook are presented in the 
final section {\bf VI}, as a discussion of the introductory points.

\section{Regge fits and determination of intercepts}

\subsection{Determination of the Pomeron and Reggeon intercepts 
from H1 data}

The diffractive
structure function $F_2^{D(3)}$, measured from DIS
events with large rapidity gaps,
can be investigated in the framework of Regge
phenomenology and expressed as a sum of two factorized 
contributions corresponding to a Pomeron and secondary Reggeon trajectories.

\begin{eqnarray}
F_2^{D(3)}(Q^2,\beta,x_{\PO})=
f_{\PO / p} (x_{\PO}) F_2^{\PO} (Q^2,\beta)
+ f_{\RO / p} (x_{\PO}) F_2^{\RO} (Q^2,\beta) \ .
\label{reggeform}
\end{eqnarray}

In this parameterisation,
$F_2^{\PO}$ can be interpreted as the Pomeron structure function  
and $F_2^{\RO}$ as an effective Reggeon structure function,
with the restriction that it 
takes into account various secondary Regge contributions which can hardly be 
separated.
The Pomeron and Reggeon fluxes are assumed to follow a Regge behaviour with  
linear
trajectories $\alpha_{\PO,\RO}(t)=\alpha_{\PO,\RO}(0)+\alpha^{'}_{\PO,\RO} t$, 
such
that

\begin{equation}
f_{{\PO} / p,{\RO} / p} (x_{\PO})= \int^{t_{min}}_{t_{cut}} 
\frac{e^{B_{{\PO},{\RO}}t}}
{x_{\PO}^{2 \alpha_{{\PO},{\RO}}(t) -1}} {\rm d} t 
\label{flux}
\end{equation}
where $|t_{min}|$ is the minimum kinematically allowed value of $|t|$ and
$t_{cut}=-1$ GeV$^2$ is the limit of the measurement. 

In the first step we determine the value
of the Pomeron and Reggeon intercepts ($\alpha_{\PO,\RO}(0)$) from 
phenomenological
fits of H1 data. 
In equation (\ref{reggeform}), the values of $F_2^{\PO,\RO}$ 
are treated as  free parameters at each 
$\beta$ and $Q^2$ point.
The values of
$\alpha_{{\PO},{\RO}}(0)$ are free parameters while
$B_{{\PO},{\RO}}$ and $\alpha^{'}_{{\PO},{\RO}}$ are taken from hadron-hadron 
data ($\alpha^{'}_{\PO}=0.26$ GeV$^{-2}$, $\alpha^{'}_{\RO}=0.90$ GeV$^{-2}$,
$B_{\PO}=4.6$ GeV$^{-2}$, $B_{\RO}=2.0$ GeV$^{-2}$), since the data are
not precise enough to determine the $t$-slopes \cite{f2d94}. 
In order to avoid the region that may be most strongly
affected by a non-zero value of 
the ratio $R$ of the longitudinal to the transverse 
cross-sections, only data with $y < 0.45$ 
are included in
the fit. 

Note that the fit with a single trajectory
does not give a good description of the data 
($\chi^2 /{\rm dof}=$ 65.5/19 with statistical errors only). 
This confirms the observations \cite{f2d94,gammap} 
that secondary trajectories in addition to the
Pomeron are required to describe diffractive $ep$ data. 
A much better fit is thus obtained 
when both a 
leading ($\PO$) and a sub-leading ($\RO$) trajectory  are considered.

Since the sub-leading exchange is poorly constrained by the data,
a second fit is performed to the H1 data in which the values of
$F_2^{\RO} (Q^2,\beta)$ are taken from a parameterisation of the 
pion structure function \cite{GRVpion}, with a single free normalization. 
We use the pion
structure function of Ref.\cite{GRVpion} to be able to extrapolate \footnote{Note 
that the Duke Owens parametrisation \cite{DO} used in 
Ref. 
\cite{f2d94} is not valid at low $Q^2$, as it has a constant value for a $Q^2$
less than 4 GeV$^2$.} our 
fits to 
lower $Q^2$. In principle, an interference term could be introduced in equation
(\ref{reggeform}), but this does not modify the $\chi^2$ of the fit.

The resulting 
value of $\alpha_{\PO}(0) = 1.20 \pm 0.09$ is in agreement with that obtained 
in Ref. \cite{f2d94}
and is significantly larger than values extracted from
soft hadronic data ($\alpha_{\PO} \sim 1.08$).
Also, we find $\alpha_{\RO}(0)=0.62 \pm 0.03$ \footnote{
Introducing interference terms reduces the value of the Reggeon
exponent to 0.5, compatible with regge phenomenology. The fact we do not put any 
interference term leads
to an effective value of the Reggeon exponent of 0.62. On the other hand,
the value of the Pomeron exponent remains constant with or without
interference.}. The use of the 
pion structure function
in our treatment is questionable as the Reggeon intercept is close
to the rho value and not to the pion one. However, the H1 
data where a proton or a neutron is tagged in the final state can be described
nicely by pion and Reggeon exchanges where the Reggeon structure function is 
also
taken
to be the pion one \cite{papfpsh1}. We thus do not expect a significant bias due 
to this assumption. 

A well-known technical problem for the determination of the Regge fit  at each 
$\beta$ and $Q^2$ point is the correct identification of the actually free 
parameters. We use a new method described in appendix {\bf A1} 
and 
perform a direct fit
of $\alpha_{\PO}(0)$ and $\alpha_{\RO}(0)$.
Indeed, this method  is designed to take into account the
decorrelations
between the parameters and to give a more reliable minimization 
procedure. Without using this method, many local minima are present in the 
$\chi^2$ which might give an incorrect value of the exponents.

This method leads to better and faster convergence properties for the
Regge fits procedure when applied to H1 data since the number of
parameters to be fitted is much smaller.
We find the following values~:
$\alpha_{\PO}(0)=1.20 \pm 0.02$ and $\alpha_{\RO}(0)=0.62 \pm 0.02$ (statistical
errors only), 
which are close to
those determined previously. We keep these
values fixed in the next sections. 

\subsection{Determination of the Pomeron and Reggeon intercepts 
from ZEUS data}

We performed the same study using ZEUS data and we find a Pomeron intercept 
$\alpha_{\PO} = 1.13 \pm 0.04$, compatible with what has been
obtained by the ZEUS collaboration \cite{zeus}. This value is lower than
the H1 value \footnote{This last value is more compatible with soft physics
if one considers a global fit to all data from $\pi p$, $K p$ and $pp$
including the 1800 GeV $p \bar{p}$ cross-section \cite{dino}.}
extracted in the previous section, however, the selection
of the diffractive sample in ZEUS is different from 
H1 \cite{zeus}. We will come back to this last point in the
following. 
The Reggeon contribution is also found to be compatible with zero
since the method of measurement (the so-called $M_X$ method) is
not sensitive to the Reggeon component \cite{zeus}.

\section{Extraction of parton distributions in the Pomeron}

\subsection{Parton distributions from H1 data}

As we have already indicated in Equation (1), 
the diffractive structure function $F_2^{D(3)}$ can be investigated  in the 
framework of Regge phenomenology.
Moreover, it has been suggested that
the $Q^2$ evolution of these structure functions may be understood in terms
of parton dynamics \cite{ingelman}, i.e. as coming from leading twist, 
perturbative QCD contributions 
where parton densities are evolved according to DGLAP equations
\cite{dglap}. 

This idea has been tested in Ref. \cite{f2d94}
with an iterative method using the CTEQ evolution code \cite{cteq}
over infinitesimal steps in $x$ space
to solve the NLO DGLAP equations \cite{dglap}. In the following we 
redo this analysis of H1 data with a new polynomial method
to solve DGLAP equations at NLO \cite{lolo}. 

We assign parton distribution functions to the Pomeron and to
the Reggeon. A simple
prescription is adopted in which the parton distributions of 
both the Pomeron
and the Reggeon are parameterised in terms of non-perturbative input
distributions at some low scale $Q_0^2= 3$ GeV$^2$. 

As we said in the last chapter, the pion structure function 
\cite{GRVpion} is assumed for
the sub-leading Reggeon trajectory
with a free global normalization to be determined by the data \footnote{We 
checked that changing
the pion structure function by some amount (20\%) does not change significantly
the parton distributions.}. 

For the Pomeron, a quark flavour singlet distribution
($z{ {S}}_{q}(z,Q^2)=u+\bar{u}+d+\bar{d}+s+\bar{s}$)
and a gluon distribution ($z{\it {G}}(z,Q^2)$) are parameterized in terms
of coefficients $C_j^{(S)}$ and $C_j^{(G)}$ at
$Q^2_0=3$ GeV$^2$ as it was done in Ref. \cite{f2d94} such that

\begin{eqnarray}
z{\it {S}}(z,Q^2=Q_0^2) &=& \left[
\sum_{j=1}^n C_j^{(S)} \cdot P_j(2z-1) \right]^2
\cdot e^{\frac{a}{z-1}} \\
z{\it {G}}(z,Q^2=Q_0^2) &=& \left[
\sum_{j=1}^n C_j^{(G)} \cdot P_j(2z-1) \right]^2
\cdot e^{\frac{a}{z-1}}
\label{gluon}
\end{eqnarray}
where $z=x_{i/I\!\!P}$ is the fractional momentum of the Pomeron carried by
the struck parton (in the following, we use indifferently $z$ or $\beta$), 
$P_j(\zeta)$ is the
$j^{th}$ member in a set of Chebyshev polynomials, which are chosen such that
$P_1=1$, $P_2=\zeta$ and $P_{j+1}(\zeta)=2\zeta P_{j}(\zeta)-P_{j-1}
(\zeta)$. 

A sum of $n=3$ orthonormal polynomials is used so that the input
distributions are free to adopt a large range of forms for a
given number of parameters.  Any bias towards a particular solution due to
the choice of the functional form of the input distribution is therefore
minimized.  

The trajectory intercepts are fixed to $\alpha_{\PO} = 1.20$ and
$\alpha_{\RO} = 0.62$ \footnote{If one fits the values
of $\alpha_{\PO}$ and $\alpha_{\RO}$, one gets the same values
as in Section {\bf II}.} (as explained in the last section). 
Only data points of H1  
with $Q^2 \ge 3$ GeV$^2$, $\beta \le 0.65$, $M_X > 2$ GeV and $y \le 0.45$
are included in the fit in order to
avoid large higher twist effects and the region that may be most strongly
affected by a non-zero value of $R$. We will discuss this point in the
following when we take into account the higher twist contribution to the
diffractive structure functions.
The fits include
161 data points for 7 parameters (3 for the sea quark density,
3 for the gluon density and 1 for the normalization of the
Reggeon contribution). Thus, we have 154 degrees of freedom. 

To get solvable evolution equations, the parton distribution functions
must approach zero as $z\rightarrow 1$. This is achieved by introducing the
exponential term with a positive value of the parameter $a$.  Unless
otherwise indicated, in the following fits $a$ is set to $0.01$ such that
this term only influences the parametrisation in the region $z>0.9$. This
term is only present to ensure the convergence and plays a role in a domain
where we do not include data. (For instance, at $\beta=$0.65, its value
is equal to 0.97.) 

The functions $z{\it{S}}$ and $z{\it{G}}$ are evolved to higher $Q^2$ using the
next-to-leading order DGLAP evolution equations. The contribution to
$F_2^{I\!\!P}(\beta,Q^2)$ from charm quarks is calculated in the fixed
flavour scheme using the photon-gluon fusion prescription given
in~\cite{ghrgrs}. The contribution from heavier quarks is neglected. 

No momentum sum rule is imposed because of the theoretical uncertainty in
specifying the normalization of the Pomeron or Reggeon fluxes 
and because it is not clear that such a sum rule is appropriate for
the parton distributions of a virtual exchange. After fitting, however, we
observed that the sum rules were fulfilled within 10\% despite
being not required in the fits. 

The resulting parton densities of the Pomeron are
presented in Fig. \ref{f1}. One possible fit quoted here as {\it fit 
1}
shows a large gluonic content. The quark contribution is much smaller 
compared to the gluon one. An other much
less favoured fit, quoted here as {\it fit 2},
has a peaked gluon distribution at high $\beta$
\footnote{ 
{\it Fit 2} shows a higher $\chi^2$
compared to {\it fit 1} (the $\chi^2$ per degree of freedom is 1.15 for {\it fit 
1} and
1.25 for {\it fit 2}). Furthermore, {\it fit 2} is quite unstable: changing
the parameters a little modifies the gluon distribution at high $z$,
and removing some data points at low or medium $\beta$ for instance destroys
this fit completely. The quark distribution is
similar for both
fits, but the gluon distribution tends to be quite 
different at high values of $z$.
This can be easily explained as no data above $z=0.65$ are included 
in the fits.
Thus there is no constraint from the data at high $z$. Therefore, 
we keep {\it fit 1} as the best fit for
H1 diffractive cross-sections.
}. It should be noticed that only the first solution is found if one imposes
in the fit that the parton distributions have no zero over the full $\beta$
and $Q^2$ range, whereas the second fit appears when this condition is removed.

The appearance of a second solution in our fit to the H1 data is
very reminiscent of the results presented in Ref.
\cite{bartels}. In fact, the
general characteristics of the two solutions obtained in the present
paper and of those found in Ref. \cite{bartels} are quite similar: 
the difference 
in the gluon content at large $z$ (or $\beta$), and the instability of the
high-gluon solution with respect to cuts in the data used for the fit.
Nevertheless,
one has to be careful in comparing the results of Ref. \cite{bartels}
with the present
ones: whereas the former ones represent fits to the diffractive
cross section $F_2^{D(3)}$, the present
analysis describes results for diffractive parton densities.
 
\begin{figure}[htb]
\vspace{0cm}
\centering
\mbox{
\psfig{file=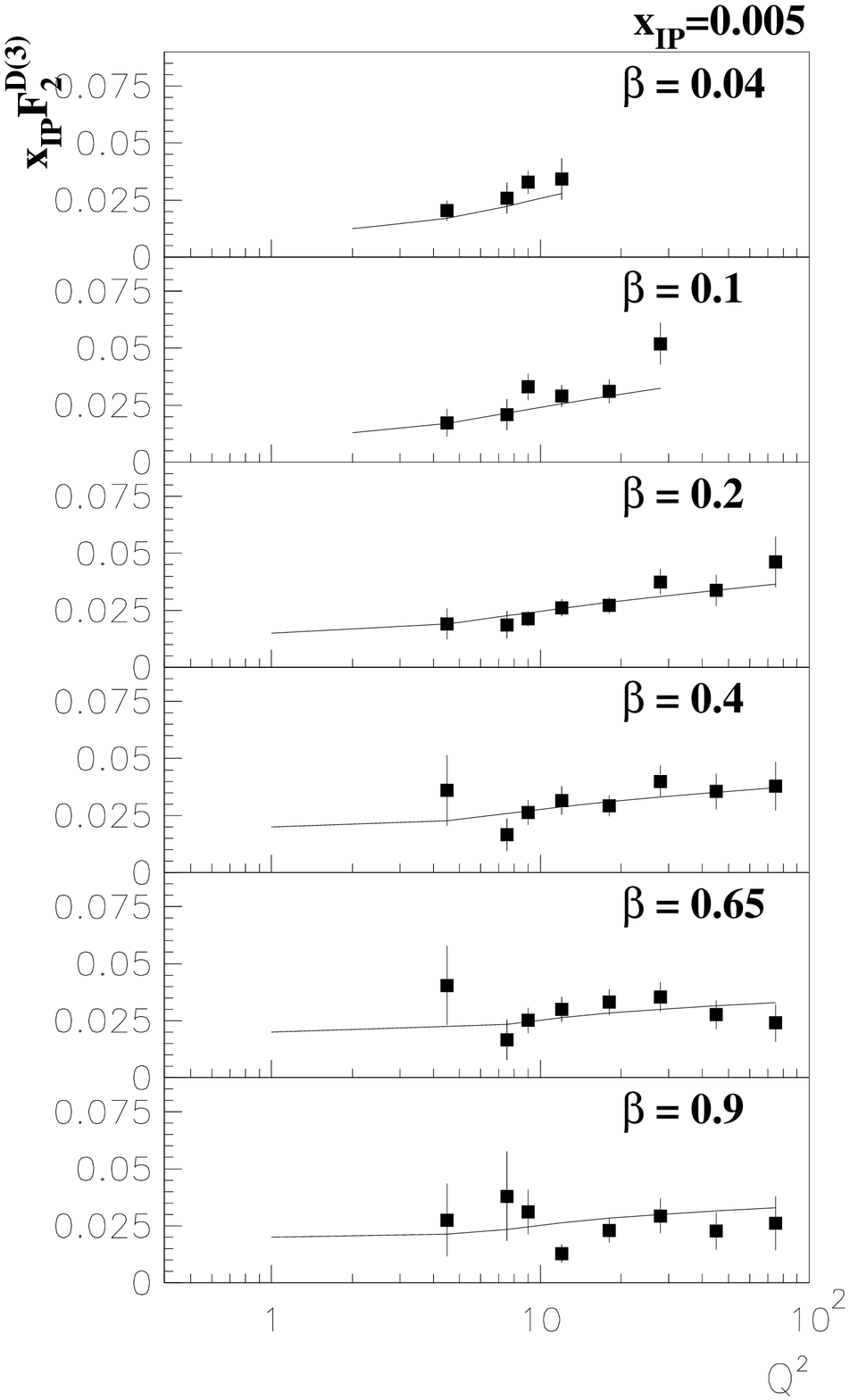,height=17cm,width=13cm}}
\vspace{0cm}
\caption{
The structure function
$x_{\PO} \cdot F_2^{D(3)}$ at $x_{\PO}=0.005$, presented
as a function of $Q^2$ in
bins of $\beta$, over the full $Q^2$ range accessed
with the H1 data.
The superimposed line represents the result
of the QCD fit to H1 data described in section IIIA.}
\label{btplot}
\end{figure}

The result of the fit ({\it fit 1}) is presented in Fig. \ref{f2} together
with the experimental values
for H1 data points; we see on this figure the good
agreement of the QCD fit with the data points, which supports the
validity of the description of the Pomeron in terms of
partons following QCD dynamics. In Fig. \ref{btplot} we also
compare the scaling violations obtained in the fit to the H1 measurement.

Let us discuss the assumptions we made in the QCD fits. We took $Q_0^2=3 \GeV^2$ 
and we noticed that the $\chi^2$ of the fit increases significantly if we
try to lower this value of $Q_0^2$ ($\chi^2/$dof=1.5 if
$Q_0^2$=1 GeV$^2$), indicating that the fit cannot be extended
to lower $Q^2$. We will come back to this
point in the following.

We performed the fits including three Chebyshev polynomials. Including one more 
polynomial
does not improve the $\chi^2$ of the fit, while two polynomials are also not
enough to get a good convergence of the fit ($\chi^2 /$dof=1.7). 

The parametrizations used in this analysis are given in formulae (3) and (4) : 
indeed, the behaviour at large $z$ is forced to go to 0 when
$z$ goes to 1
by the term $exp(\frac{a}{z-1})$ which includes an essential
singularity at $z=1$. We tried to perform the analysis with $a=0$
imposing the behaviour at large $z$ with a term in 
$(1-z)^\alpha$. With these assumptions, we did not get any fits
with a good $\chi^2$ ($\chi^2$/dof $>$2). The form of the
parametrization for the parton distribution at large $z$ 
plays a non-negligible role in the fit procedure to ensure a quick
convergence at high $z$. However, this term is large in a region
where no data are included in the fit. On a more physical ground, it is interesting to note that this behaviour is
in 
contrast with the one for proton (or even pion, cf. Drell-Yan analysis) 
structure functions for which the behaviour of the distribution tail is seen in 
a rather sizable  region of $z.$

The positive scaling violations of diffractive
structure functions lead to an important
gluon content $zG(z)$ in the Pomeron
(even at large $z$), which is what we get from QCD fits
(see Fig. \ref{btplot}).

Moreover, with this gluon distribution, 
larger scaling violations at low $\beta$ are naturally associated
with larger values of gluon splitting functions when $\beta$ gets lower.
Now, it is also clear that, due to large measurement uncertainties,
scaling violations and thus the determination
of $\alpha_s  zG$ are not very precise. The uncertainty on the value
of $\alpha_s$ is negligible in this analysis, and we can get direct access
to the uncertainty on the gluon density (determined from QCD fits). We get
from the investigation of the scaling violations 
$$
\frac{\delta G}{G} \simeq 25 \% \ .
$$

We believe that this (even not yet precise) error determination is a relevant 
piece of information, showing that the gluon density is not yet very
well constrained. 

In conclusion, the gluon content of the Pomeron is confirmed to
be large (w.r.t. \cite{f2d94}). The flat gluon solution is favoured, and 
another result ({\it fit 2}) appears if we allow that
the parton distributions can possibly vanish in the kinematical range.
Moreover this second solution is disfavoured due to
its worse $\chi^2$ and instability and thus, we will not use this second solution
in the following.
The accuracy of the gluon determination is still quite poor.
This provides a very nice perspective for diffractive analyses of more
recent data where more accurate results will lead to a more precise QCD
analysis.

\ffig{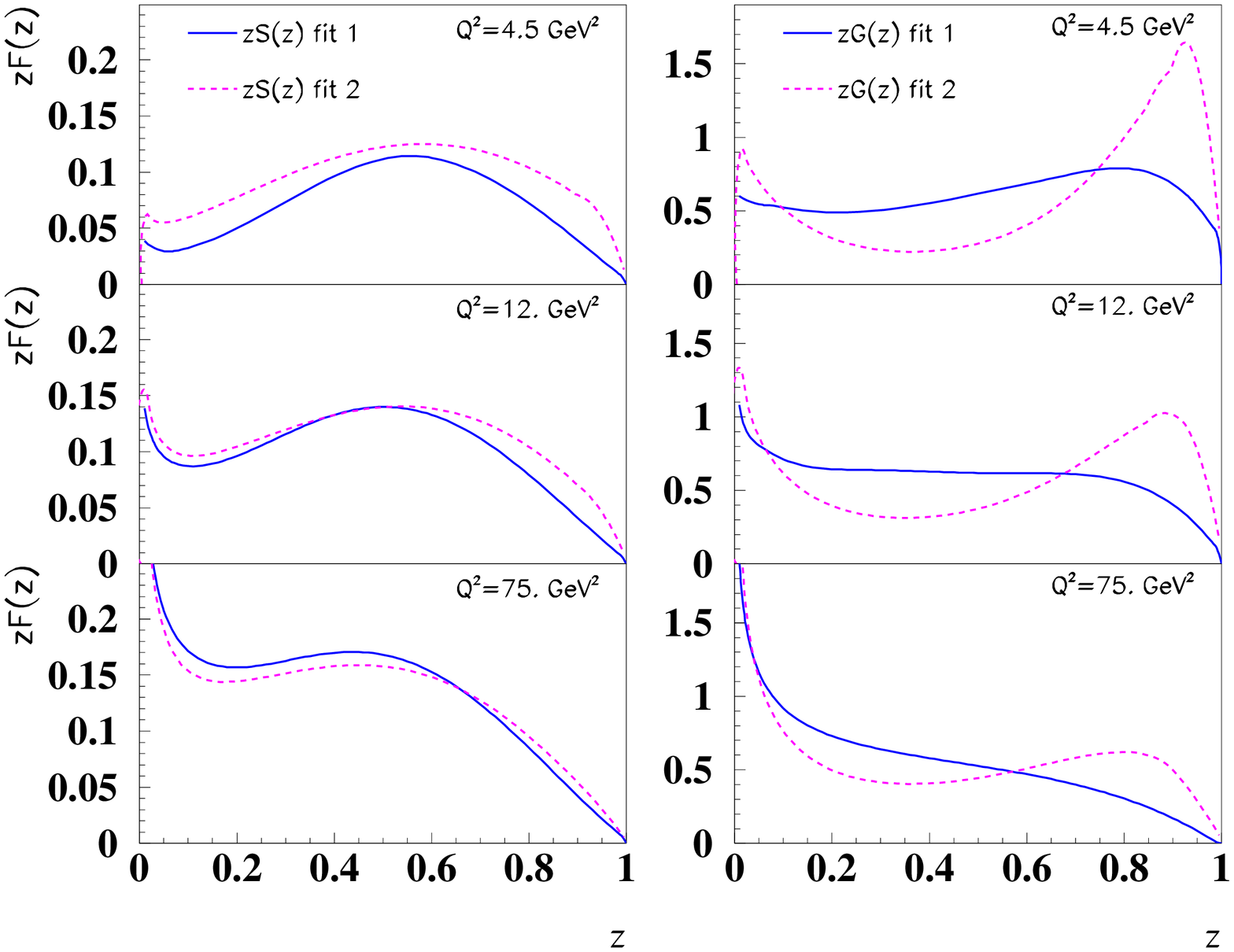}{140 mm}{
Quark flavour singlet ($zS$, left) and gluon ($zG$, right) distributions
of the Pomeron deduced as a function of $z$, the fractional momentum of the
Pomeron carried by the struck parton, from the fit
on H1 data points with $Q^2 \ge 3$ GeV$^2$. 
The solid (dashed) curve shows {\it fit 1} ({\it fit 2}) as discussed in the
text
($\chi^2/dof = 177.1/154 = 1.15$ for {\it fit 1}
and $\chi^2/dof = 192.5/154 = 1.25$ for {\it fit 2} with 
statistical errors only). The parton densities are normalised to represent 
$\xpom$ times the true parton densities multiplied by the flux factor at
$\xpom = 0.005$ (this will always be the case in all the following figures).}
{f1}

\ffig{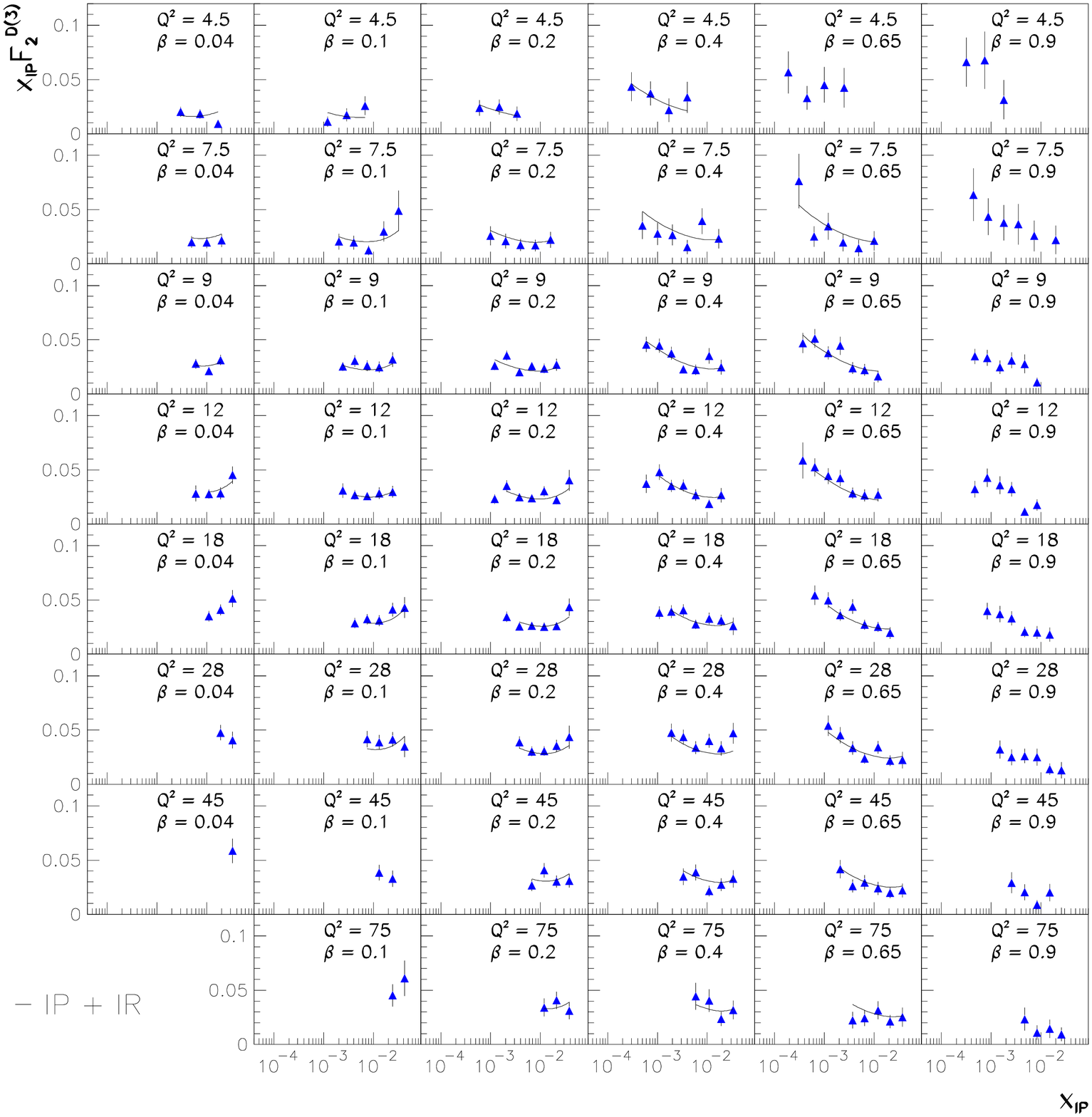}{160 mm}{
The H1 data points for $\xpom F_2^{D(3)}$
are shown with the result
of the QCD fit ({\it fit 1}) described in the text; the result of the
fit is shown only in bins
included in the minimization procedure.}
{f2}

\subsection{Parton distributions from ZEUS data}

We have redone this QCD analysis with ZEUS 
diffractive cross-section measurements \cite{zeus}
applying the same cuts.
Note that for the ZEUS fits, there are only 6 parameters
since the method of measurement is not sensitive to the
Reggeon contribution \cite{zeus}. 
The Pomeron intercept is fixed to $\alpha_{\PO} = 1.13$ as
determined from Regge fits on ZEUS data \cite{zeus}.

The resulting
parton distributions are presented in Fig. \ref{fh1zeus}
together with the H1 results.  
As the Pomeron intercepts are different for H1 and ZEUS data,
the parton densities are
normalized to represent $x_{\PO}$ times the true parton densities
multiplied by the flux factor at $x_{\PO}=0.005$.

We notice some large discrepancies,
especially in the gluon density \footnote{Note that such different gluon 
densities can lead
to similar values of the structure functions $F_2^{D(3)}$ (see Fig.
\ref{fh1zeus}) because the diffractive structure function is more sensitive
to the quark density (the photon couples directly to quark). This will however lead to differences in the
charm structure function as it is much more sensitive to the gluon
structure function.}. The gluon density derived
from ZEUS measurements is lower that the H1
fit results. The data are compared directly in Fig. 6 where we rescaled the H1
data to the ZEUS bins (the rescaling is not very model-dependent since the
shifts in $Q^2$ and $\beta$ between both experiments are quite small, and we 
performed the rescaling using models in Ref. \cite{bartels} \cite{robi},
which lead to the same result). In addition to this treatment of data, about 10\%
(30 \%) of the cross-section should be subtracted to take into account
proton dissociation contribution in the H1 (ZEUS) cross-section measurement.
From Fig. 6, we notice that the main
differences are located in the highest $Q^2$ bins ($Q^2 \sim 60.$ GeV$^2$) after
this additional correction and could be due to the different methods
used to do this measurement (rapidity gap method or so-called $M_X$ method
\cite{f2d94,zeus}).
However, there are only 30 ZEUS data points 
entering in the QCD analysis compared to 161 for H1 analysis,
the experimental uncertainties being of the same order.
Thus, the statistical power of the ZEUS fits is considerably
lower. We can estimate the statistical uncertainty on the gluon
distribution to be about 50\% for ZEUS data  (to be compared with 25\% for
H1 data). The differences in the quark densities are much smaller taking
into account the fact that there is no ZEUS data at $Q^2=4.5$ GeV$^2$ and
low $\beta$.
Differences in the parton distributions are clearly due to 
small differences in the data (see Fig. \ref{fh1zeus}). We see differences
in the $Q^2$=60 GeV$^2$, $\beta$=0.7, or $Q^2$=8 GeV$^2$, $\beta$=0.2 bins.

As the fit results are quite different for both experiments, we choose 
to make predictions for the longitudinal and charm structure functions for
both sets of parameters.

\ffig{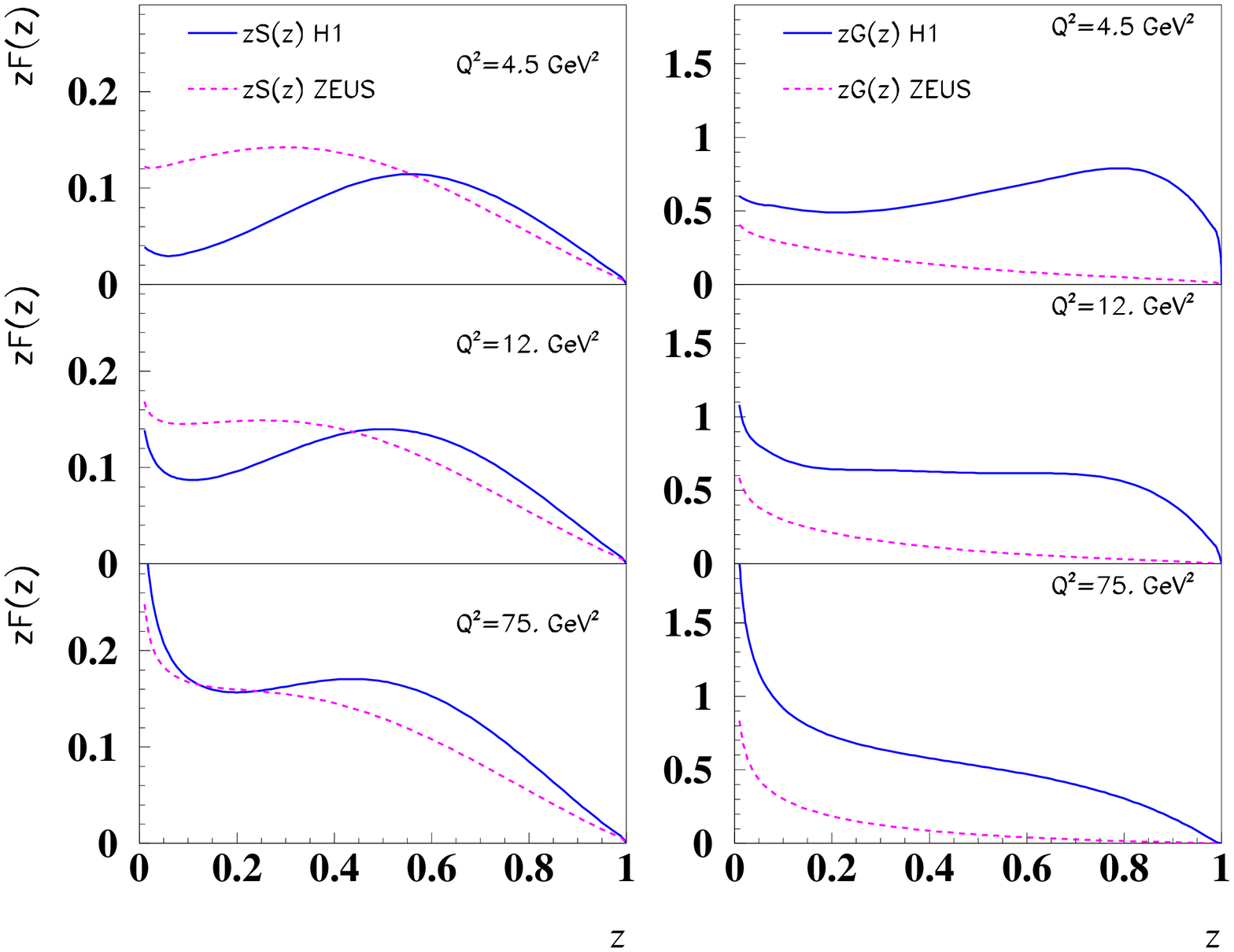}{140 mm}
{
Quark flavour singlet ($zS$, left) and gluon ($zG$, right) distributions
of the Pomeron derived from 
H1 diffractive data ($\chi^2/dof = 177.1/154 = 1.15$,
full lines) and 
ZEUS diffractive data ($\chi^2/dof = 29.3/(30-6) = 1.22$,
dashed lines).}
{fh1zeus}

\ffig{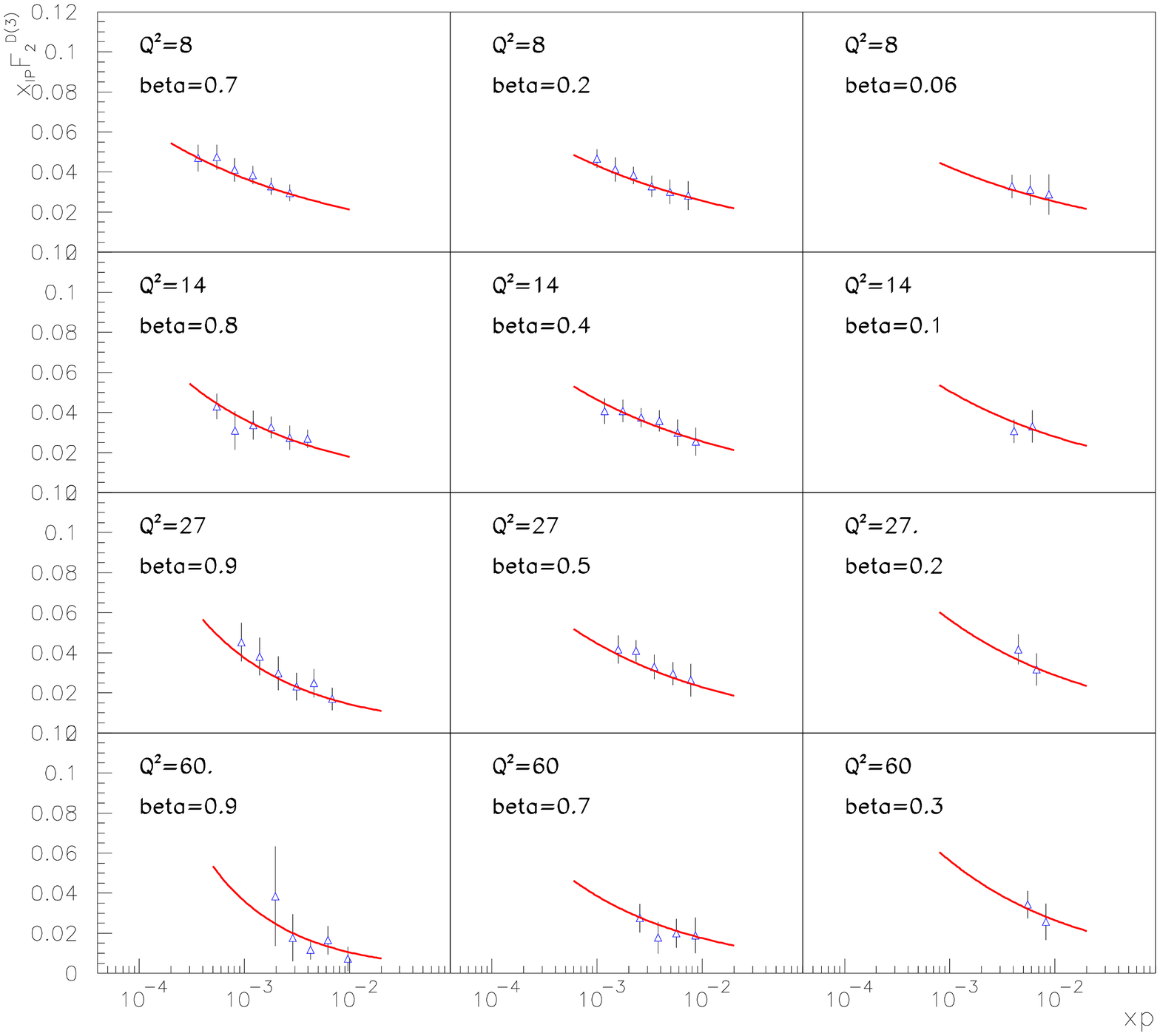}{110 mm}
{The ZEUS data points on $\xpom F_2^{D(3)}$ 
are shown with the result
of the QCD fit described in the text.}
{zeus}

\vspace{-6cm}

\ffig{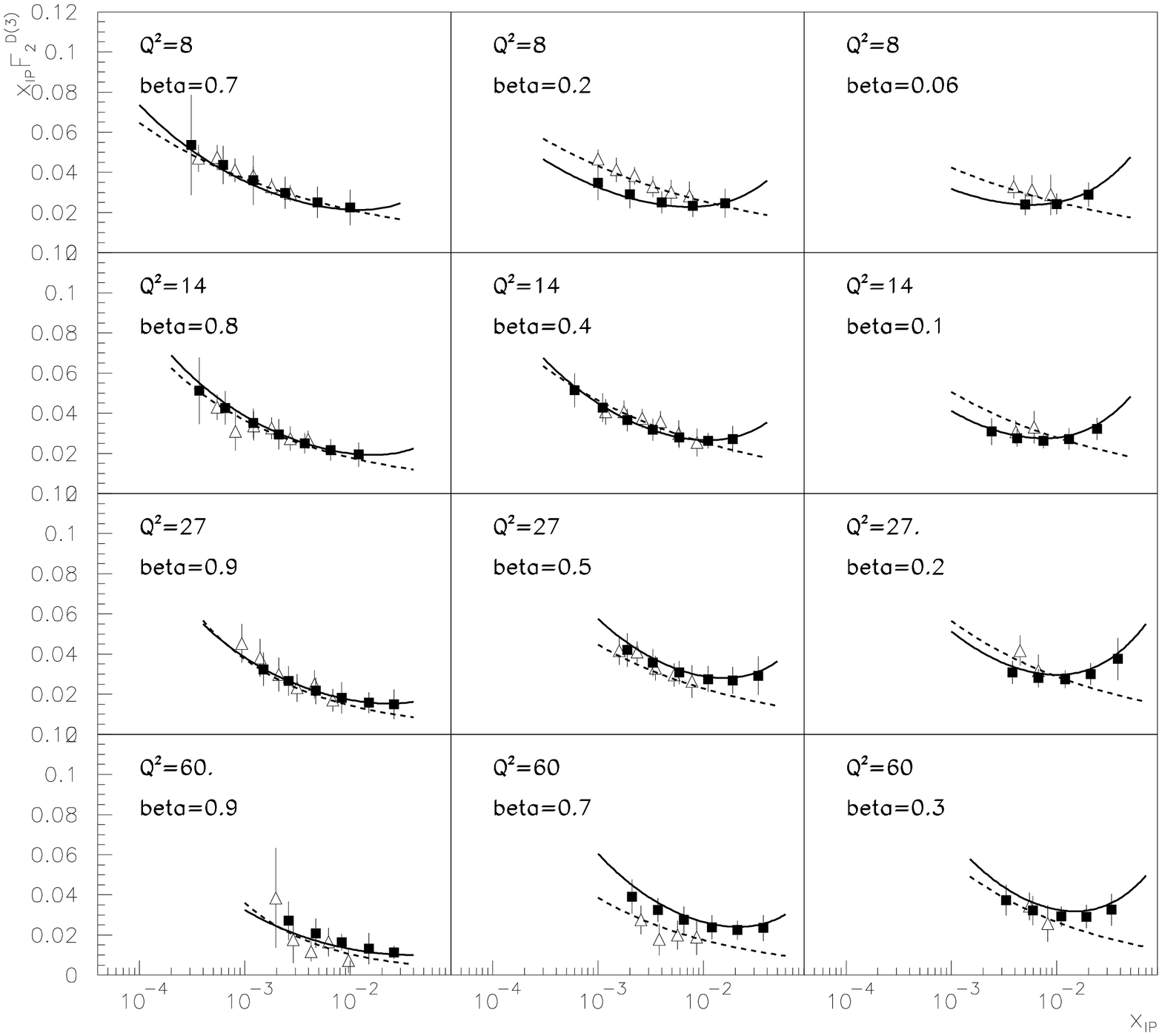}{110 mm}
{Comparison of H1 and ZEUS data on $\xpom F_2^{D(3)}$
(respectively black squares and open triangles) with the result of the QCD fits: 
full line for the H1 QCD fit result
and dashed line for
ZEUS. Note that the H1 points have been rebinned to ZEUS $Q^2$ and $\beta$
values.}
{zeush1}

\section{global fits}

\subsection{Higher-twists and global fits}

The H1 fit ({\it fit 1}) is our reference and we attempt to 
refine it by extending the number of data points included
into the fit procedure. The problem is that the cuts
described above are essential to avoid the experimental domain
where non-perturbative power
contributions could be large and we cannot suppress these cuts
without subtracting in another way these power contributions
(or higher-twist terms). However, some parametrizations
for higher-twist contributions have been proposed \cite{bartels,robi}
after a dedicated analysis of H1 data.

We use the results of Ref. \cite{bartels,robi}
to subtract the higher twist component from the $F_2^{D(3)}$ 
measurements at high $\beta$ and
we redo the QCD analysis with all data points
included, which gives
224 data points for H1. 

The 
resulting parton distribution are presented in Fig. \ref{fh1h1m} for the fit
to the H1 data using the higher twist determination of Ref. 
\cite{bartels} \footnote{We checked that using another higher twist
parametrisation from Ref. \cite{robi} which uses a BFKL-dipole type 
approach \cite{bfkl,dipole} does not change our results.}.
We notice the perfect agreement between parton distributions
extracted with this procedure and with
our former analysis described above. However, the statistical power
of the new procedure is better due to the larger number of 
degrees of freedom. We also note that the $\chi^2$/dof is 
smaller when we perform the fit after subtracting the higher twist component
(1.05 compared to 1.15, see Fig. 7).
This implies that a small residual
higher twist component is still present near $\beta$=0.65 which 
influences the quality of the fit, and thus, it is quite dangerous to cut 
directly in the data. 

As we now include all data in the fit, the assumption we made on the $a$ 
parameter
of Equation (\ref{gluon}) may play a role. We checked as before that the 
convergence
is not possible if we include a polynomial instead of an exponential form
(the $\chi^2$ per degree of freedom increases up to 2.).
However, the value of $a$ is not fundamental as another value ($a=10^{-3}$) 
also leads
to a good fit. 

The parton distributions extracted with the global procedure
are now our reference distributions. The parameters of the gluon and quark
densities are given in appendix A2.

From the ZEUS data, we also get comparable results with or without higher
twist corrections,
leading to
the same parton distributions as shown in Fig. \ref{fh1zeus}. 

\vspace{-1cm}

\ffig{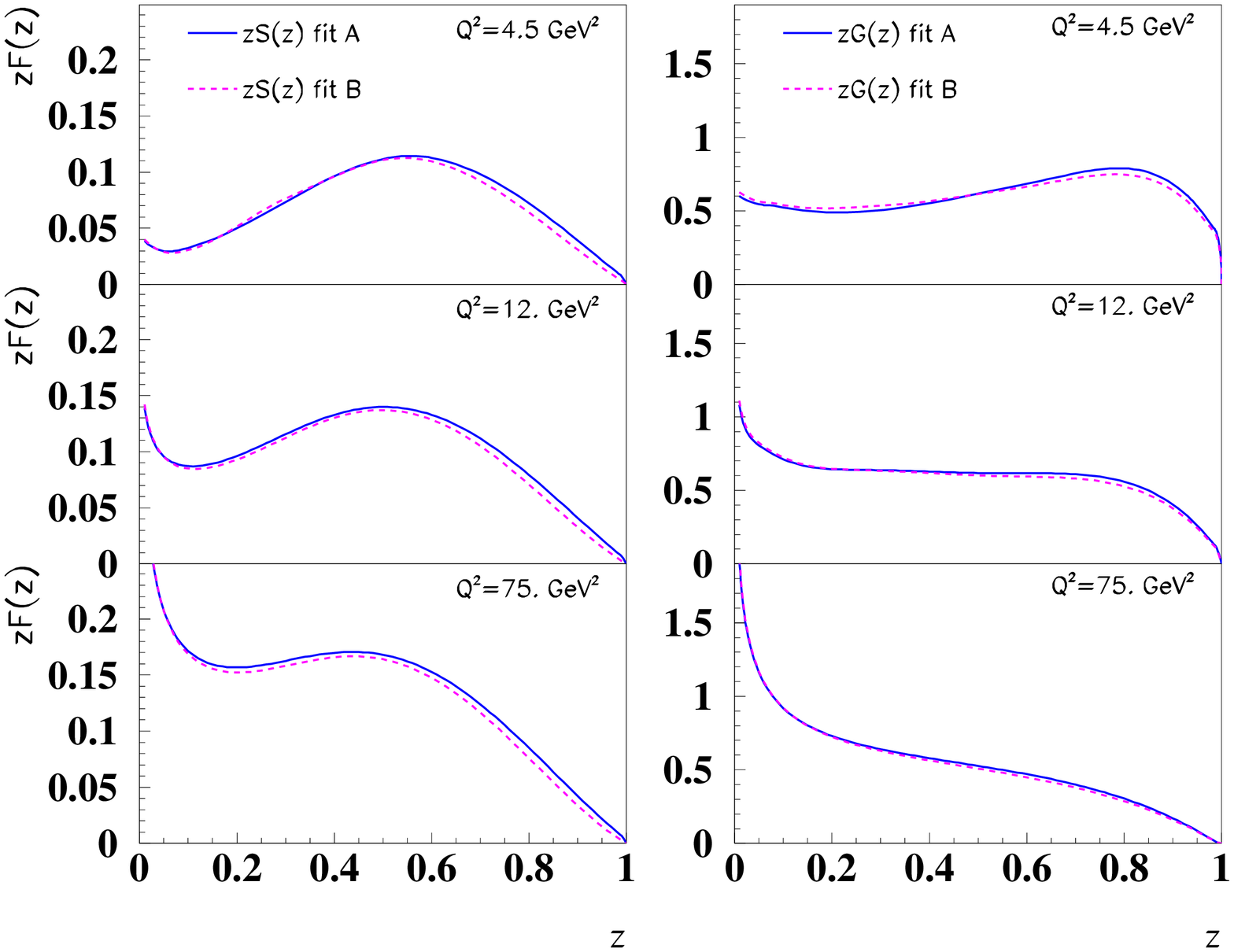}{95 mm}
{Parton distributions extracted from
H1 data with the standard QCD fits
($\chi^2/dof = 177.1/154 = 1.15$, full lines, {\it fit A}) compared to
distributions extracted from the {\it global fit} analysis using all
H1 measurements
($\chi^2/dof = 228.4/(224-6) = 1.05$, dashed lines, {\it fit B}).}
{fh1h1m}

\subsection{Comparison of the gluon density in the proton and in the Pomeron}

As can be seen in Fig. \ref{fh1h1gf2}, the gluon density derived
from our global QCD fits of H1 diffractive measurements is 
very different from the one derived from QCD fits of $F_2$ 
\cite{H1f2}. However, since the uncertainty of our determination
is large ($\frac{\delta G}{G}\ge 25 \%$ 
as we have mentioned above), one can
wonder whether $F_2^{D(3)}$ could be described using the same gluon
density as for the proton total structure function $F_2$. 

We have taken the gluon density of Ref. \cite{H1f2} (we simply identify Bjorken-$x$ with
$\beta$ from this parametrisation) 
and we have redone the QCD analysis of $F_2^{D(3)}$ H1 data
by fitting only the sea quark distribution using directly
the gluon density inside the proton.
As explained above we have also used the prescription of Ref. \cite{bartels}
to subtract the higher-twist contribution from diffractive
measurements, which gives 224 data points and 3 parameters in the fit.
This procedure leads to $\chi^2/dof = 304.2/(224-3)=1.37$
which is worse than our reference fit for diffractive data
($\chi^2/dof = 228.4/(224-6) = 1.05$).
This result suggests that the gluon density
of the Pomeron is quite different from the gluon density of the proton
\footnote{Note that we have also redone this procedure with a free normalization
for the gluon distribution. However, this 
normalization is not changed during the fitting procedure. Thus we
get the same results as before.
}. The difference is smaller between  
the result of the ZEUS fit and the gluon density inside the proton
(in the case of the ZEUS fit, the gluon density in
the Pomeron is found to be smaller than in the H1 case). Both H1 and ZEUS 
gluon densities in the Pomeron are however incompatible with the gluon
in the proton at high $x$ or $\beta$.

The fit results compared to the data show large discrepencies at low $z$
in all $Q^2$ bins. 
It is impossible to describe diffractive data at low $\beta$ using the same
gluon density as in the proton. 

\ffig{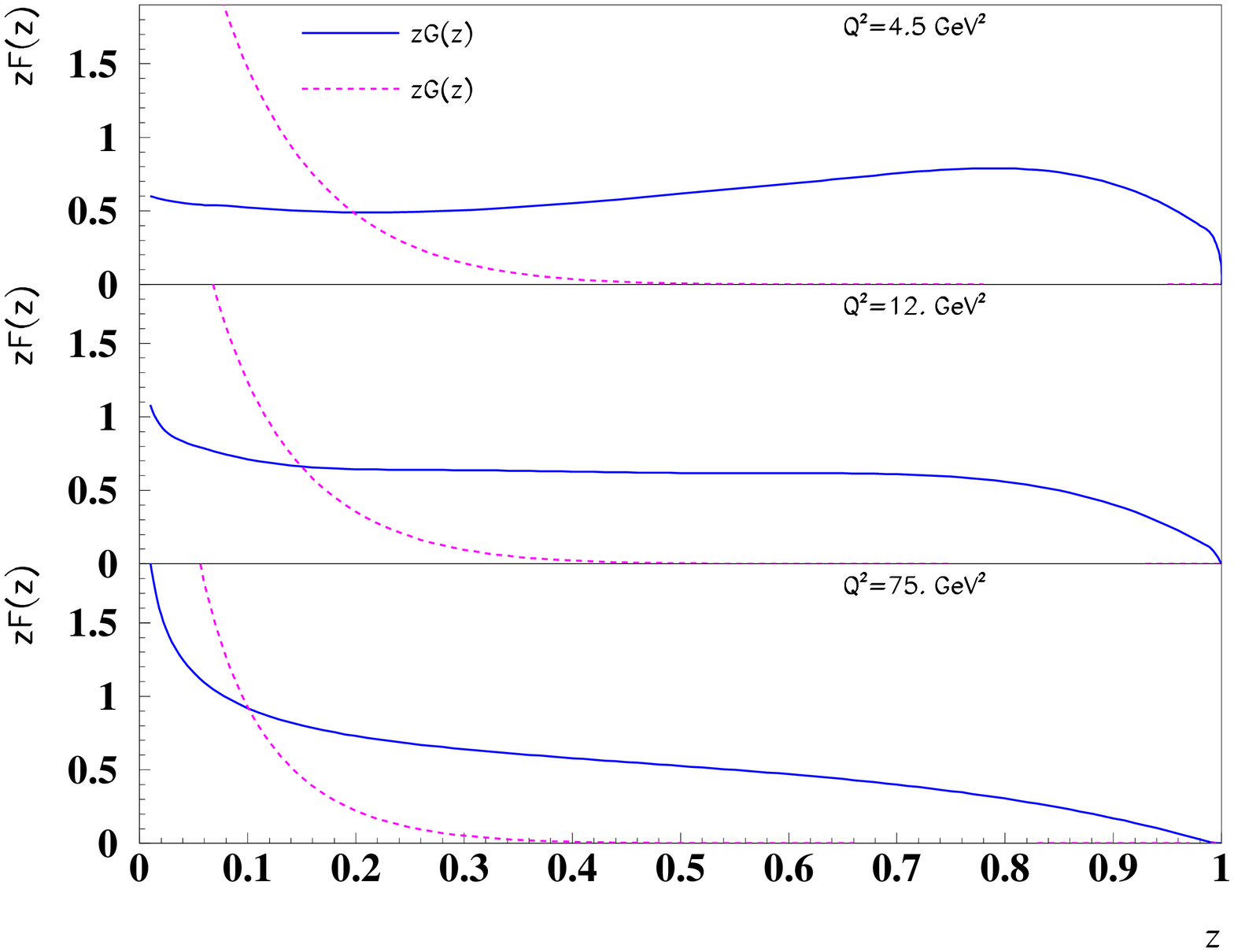}{100 mm}
{Gluon density
fitted from diffractive data (global fits, full lines) 
compared with 
the gluon density extracted from $F_2$ QCD fits
(dashed lines).}
{fh1h1gf2}

\subsection{Fit extrapolations in $Q^2$}

We now extrapolate our QCD fits using H1 and ZEUS data to low and high
$Q^2$ to look for a possible kinematical domain where the differences 
in the H1 and ZEUS parton distributions can lead to differences in
$F_2^D$. 

As we mentioned before in section IIIA, it is difficult to lower the starting scale 
$Q_0^2$ to get a prediction for $F_2^{D(3)}$ at low $Q^2$ because the quality
of the fit is much worse. Thus, we perform a low $Q^2$ backward 
evolution using $Q_0^2$=3 GeV$^2$.  
On the other hand, it has been recently noted by the
H1 and ZEUS collaborations \cite{chrdis99} that the H1 1995 
and ZEUS 1995 preliminary data at low
$Q^2$ cannot be described using our perturbative QCD approaches, which might
be an indication that perturbative QCD or a DGLAP fit to $F_2^D$ 
is not valid in this region of phase space without saturation effects
\cite{golec}.

At high $Q^2$ (see Fig. \ref{highq2}) the differences between both 
extrapolations remain small as was the case at medium $Q^2$, except at high 
$\xpom$, which is an artefact of the absence of Reggeon
contribution in the ZEUS fit.

The fact that no large difference is found between both extrapolations
whereas we have large differences in the gluon density shows that 
we need other measurements more sensitive
to the gluon density, like the charm component to $F_2^D$, or some direct 
final state studies (jet production, multiplicity studies \cite{finalstate}).


\vspace{-3cm}

\ffig{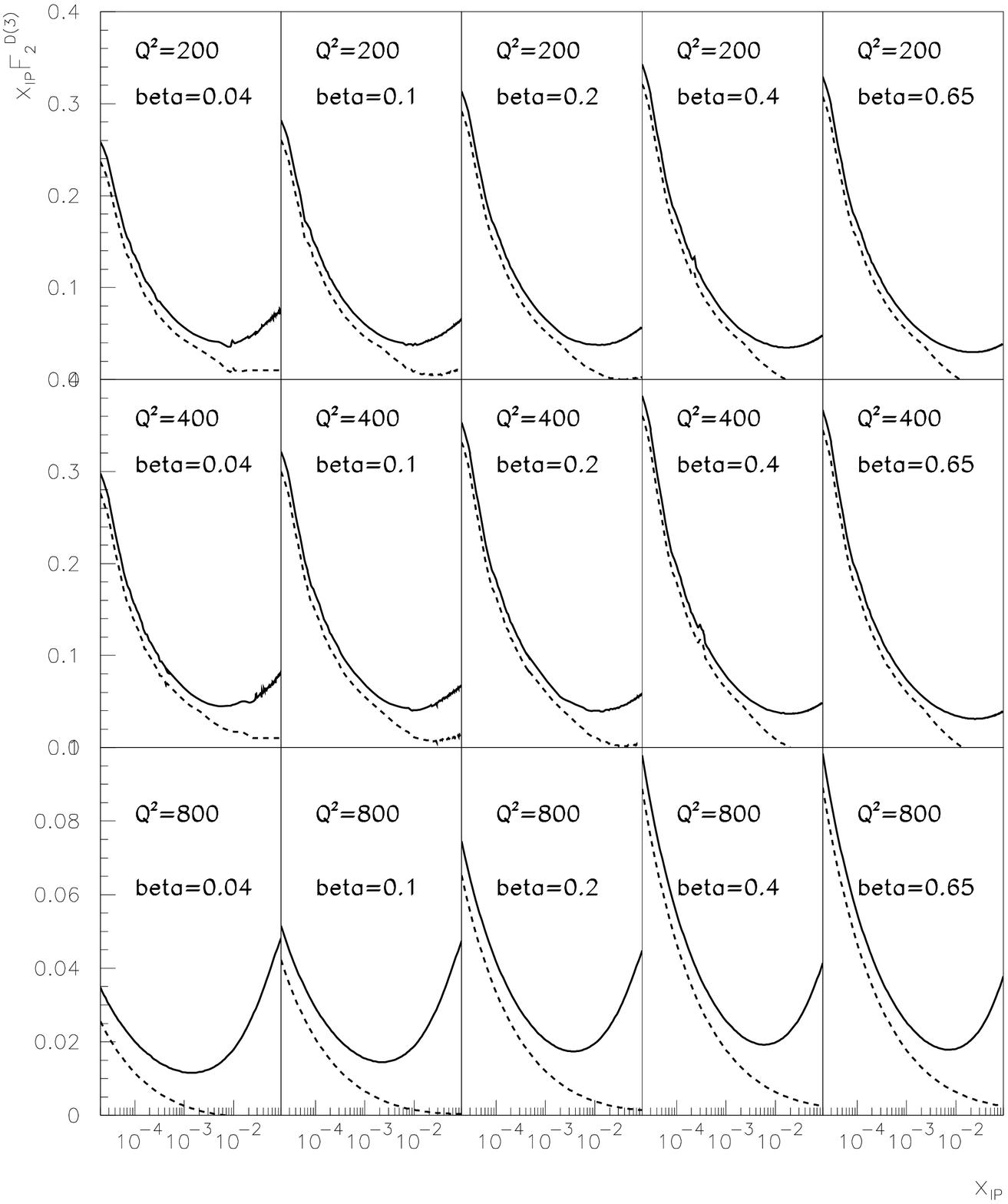}{140 mm}
{High $Q^2$ extrapolation of the fit results (H1: full line, ZEUS: dashed
line).}
{highq2}
\subsection{Charm contribution and longitudinal diffractive 
structure functions}

From global QCD fits of the diffractive structure function
$F_2^{D(3)}$, we have derived reference parton distributions
(see Fig. \ref{fh1h1m}). With these distributions, we
can now give some predictions for the longitudinal and charm structure
functions in diffraction.

We present values for the longitudinal structure function
in Fig. \ref{flh1}, which is the sum of two components, namely
the perturbative part coming from the QCD fit
and the higher-twist component.
The higher-twist component of the longitudinal structure function
becomes dominant at high value of $\beta$
whereas the low $\beta$ domain is dominated by the perturbative part
of $F_L$ deduced from QCD fits.
We can also evaluate values
of $R=\frac{\sigma_L}{\sigma_T}$ in diffraction, which is illustrated
in Fig. \ref{rh1}. Here again, the high $\beta$ domain represents
the higher-twist component. 

We note that the differences between the H1 and ZEUS fit results are quite
small in $F_L$. However, a measurement of $F_L^D$ or $R$ would be quite 
interesting
to constrain the higher twist component at high $\beta$. The $R$ values are
more different between H1 and ZEUS predictions, but the error bars on these
extrapolations are quite large.

The charm predictions are shown in Fig. \ref{fch1}, \ref{rch1}. We note
the large differences between the ZEUS and H1 fit results which shows the
importance of this direct measurement, which is directly sensitive to the 
gluon density. The charm contribution to $F_2^D$ is expected to be larger than
for the proton structure function (about 40\% at $\beta \sim$0.1,
$Q^2 \sim$ 10 GeV$^2$) if we consider the results of the H1 fit, whereas the 
results of the ZEUS fit are more similar to the proton values
\footnote{Preliminary H1 and ZEUS measurements \cite{charm} show that the H1
(resp. ZEUS) charm cross-section is closer to the results of the
predictions coming from the ZEUS (resp. H1) QCD fit. More precise data are
needed to solve these discrepencies.}.

\ffig{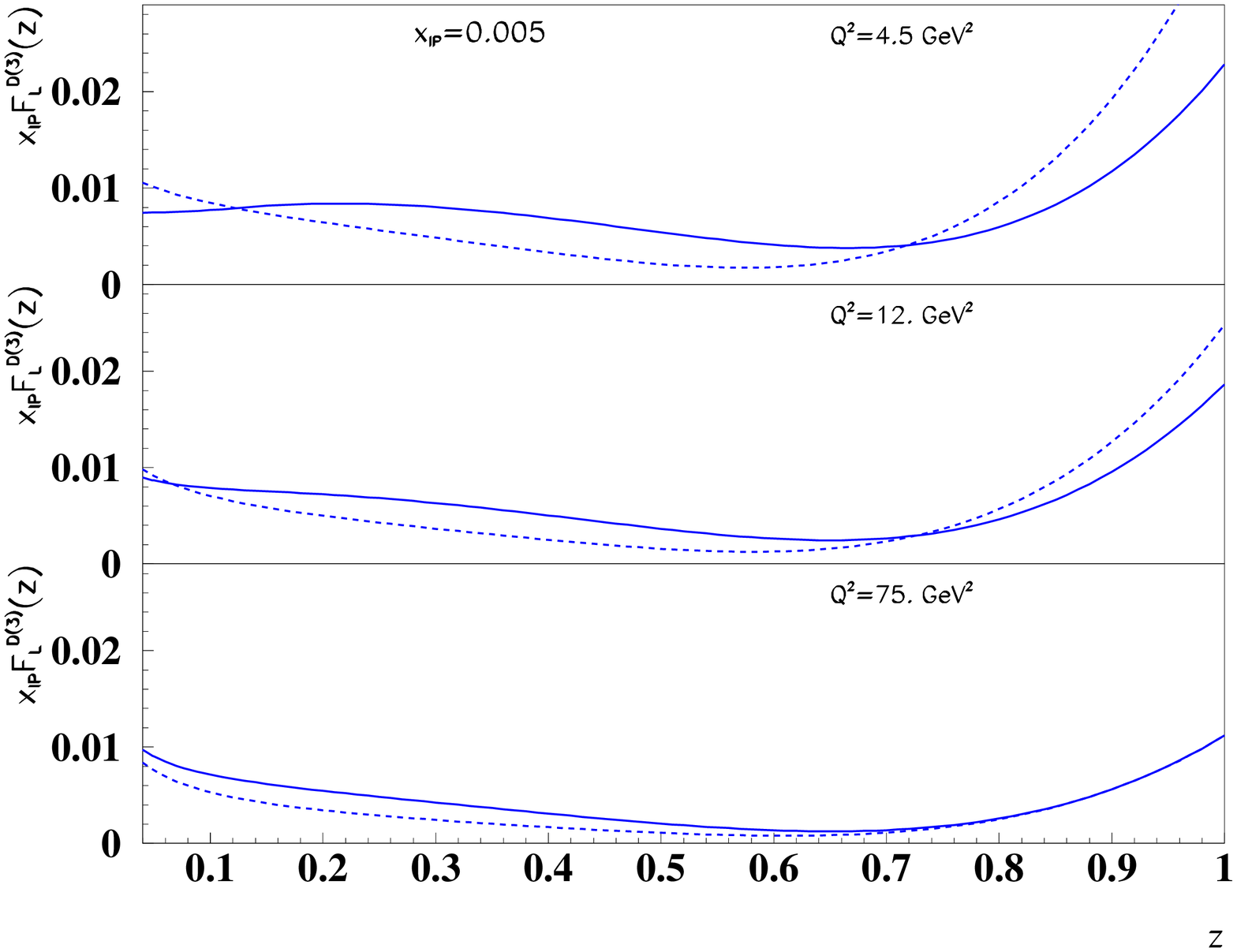}{100 mm}
{Prediction for the longitudinal structure function
$x_{\PO} \cdot F_L^{D(3)}$ (at fixed $x_{\PO}=0.005$) 
presented as a function
of $z=\beta$ for different values of $Q^2$ (H1: full line, ZEUS: dashed line).}
{flh1}

\ffig{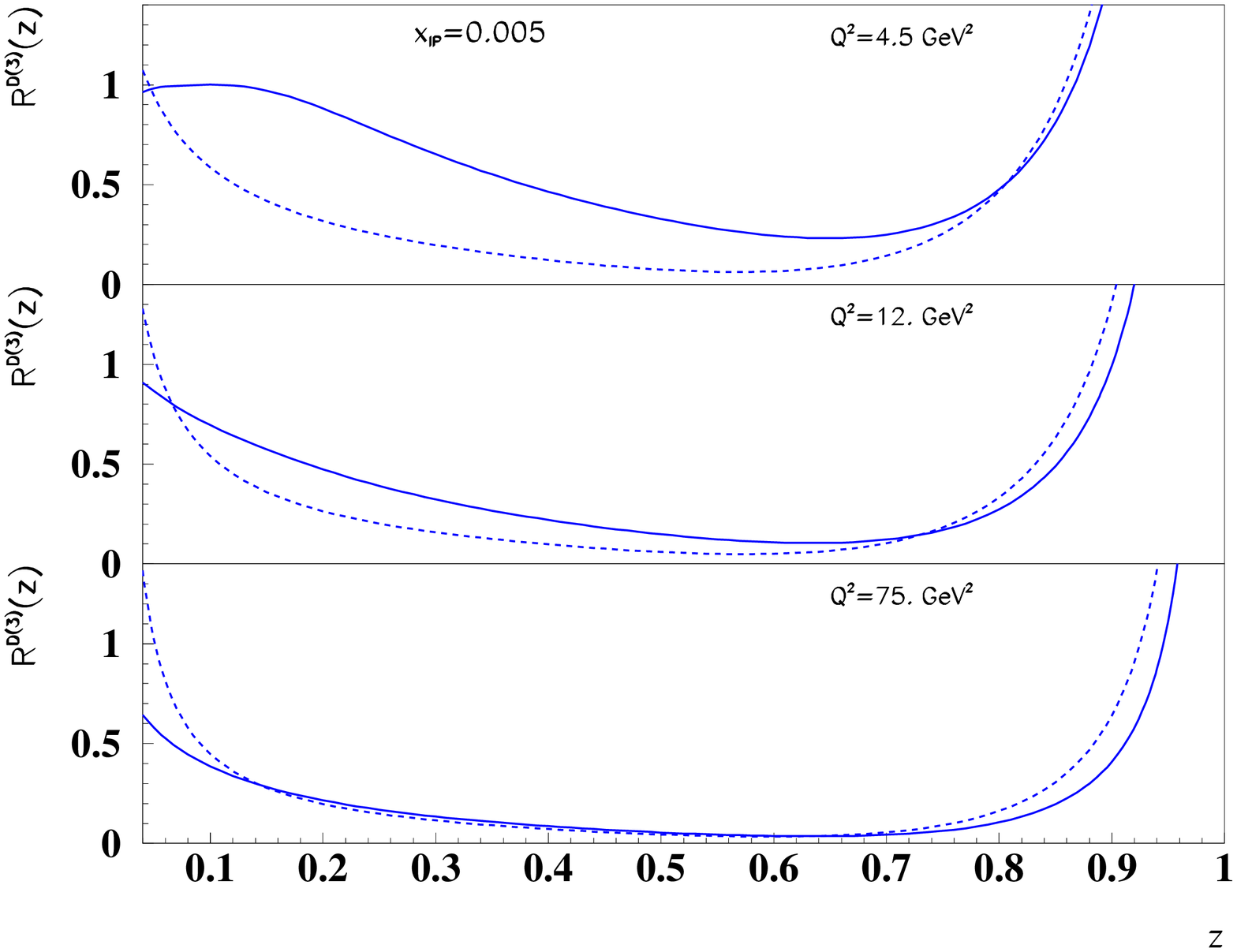}{100 mm}
{Prediction for $R^{D(3)}$ presented as a function
of $z=\beta$ for different values of $Q^2$ (H1: full line, ZEUS: dashed line).}
{rh1}

\ffig{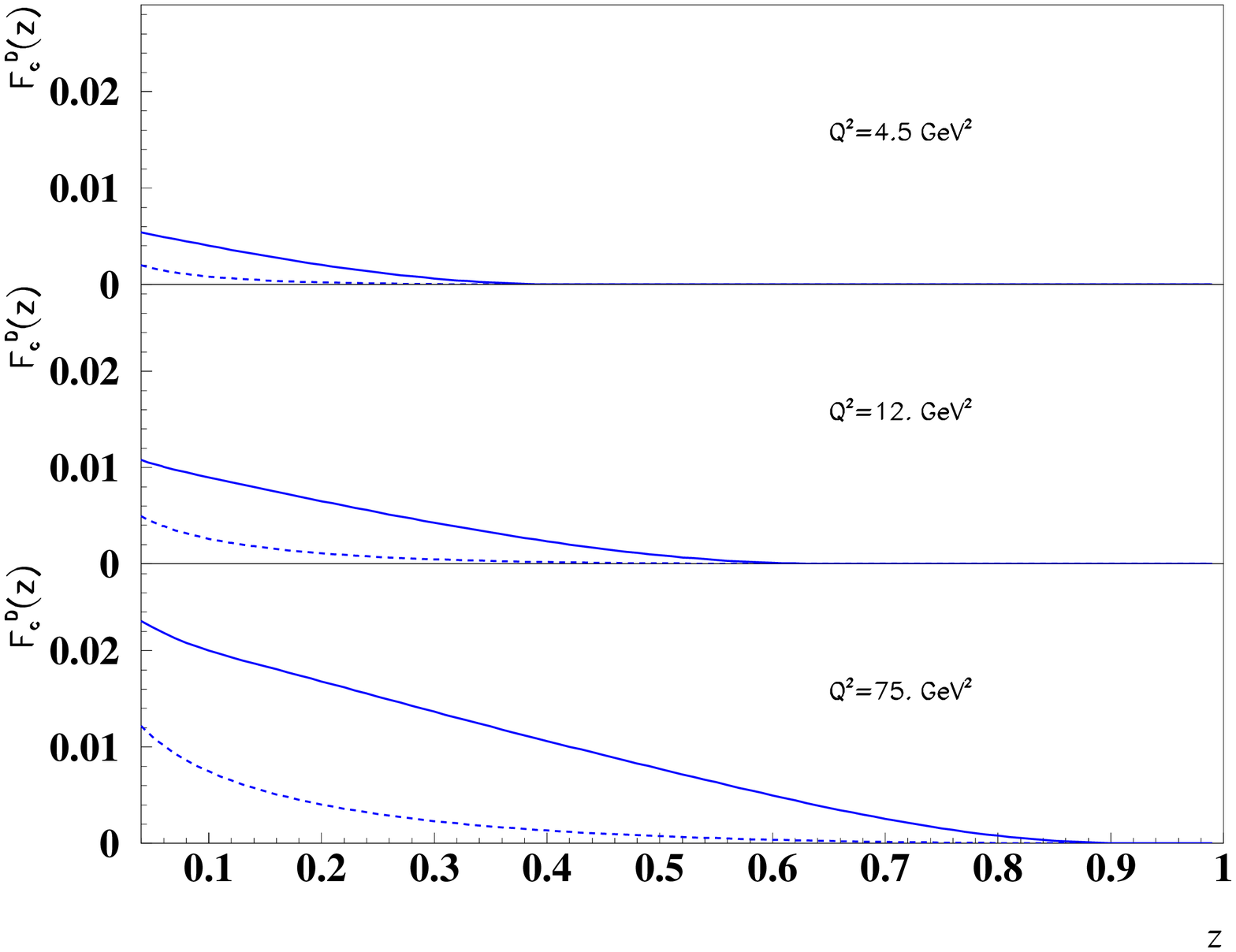}{100 mm}
{Prediction for the charm structure function
$F_c^{D}$ presented as a function
of $z=\beta$ for different values of $Q^2$ (H1: full line, ZEUS: dashed line).}
{fch1}

\ffig{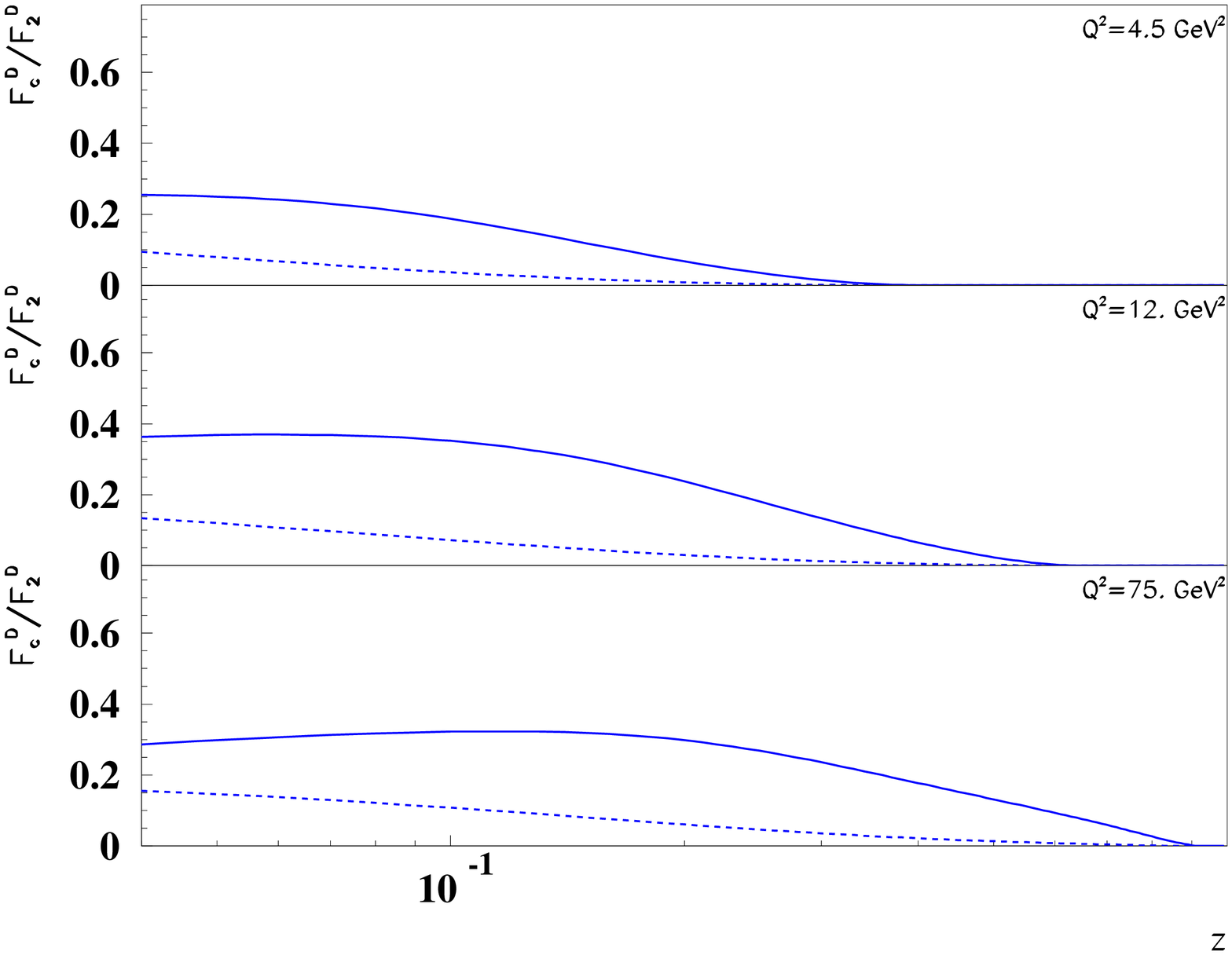}{100 mm}
{Prediction for the fraction of charm structure function
$F_2^{D (c \bar{c})}/F_2^{D}$ presented as a function
of $z=\beta$ for different values of $Q^2$ (H1: full line, ZEUS: dashed line).}
{rch1}

\section{Extrapolation to Tevatron and comparison with CDF data}
The QCD fits we  obtained from HERA data allow us to make direct
comparisons
for measurements at the Tevatron. It is quite interesting to be able to  
directly test factorization breaking between HERA and the Tevatron using 
the measurements performed at both accelerators. We thus compare the
extrapolations of the H1 and ZEUS QCD fits to the recent CDF diffractive
jet cross-section
measurement \cite{cdf}. The result is given in Fig. \ref{cdf}.
We note a large discrepancy both in shape and normalization between H1
predictions and CDF data, clearly showing factorization breaking. However,
the ZEUS fits are more compatible in normalization with the CDF measurement
even if the shape is not described properly. We know 
that the 
ZEUS gluon density is between 2 and 3 times smaller than the H1 one. The 
predictions for the Tevatron  are thus expected to be very much different by
a factor 2 to 3 between H1 and ZEUS since they
correspond to single Pomeron exchange. If the large statistical and systematic
uncertainties on the gluon density are taken into account (about 50\%
for ZEUS, 25\% for H1), ZEUS data are compatible with  
factorization 
at low $\beta$. One should, however, question whether one has the right to extrapolate
ZEUS results without introducing an additional Reggeon component. Namely
the CDF measurement is in a region in $\xpom$ where the Reggeon contribution
is important, in contrary to the ZEUS measurement. The error bar on the ZEUS fit
extrapolation is thus enhanced due to this uncertainty. Concerning the extrapolation
of the H1 fit and the comparison with CDF data, one must notice that the CDF data
lie primilarly at low values of $\beta$ where there are few H1 data points. Moreover, in this
kinematical domain, the H1 data have the tendency to lie more in the high $\xpom$
region where the Reggeon contribution is important and not well constrained by the 
fit. Thus, the extrapolation to the CDF domain suffers from large uncertainties.
A combined fit using CDF data to constrain the low $\beta$ region and the
HERA data to constrain the high $\beta$ domain would thus be of great interest.

More precise data from HERA and more detailed comparisons
with Tevatron are thus needed to study precisely factorization breaking
between both experiments. The discussion of eventual higher twist contributions 
is clearly also valuable in order to reach conclusions on the shape at larger $\beta.$
It is thus important to get an accurate measurement of the gluon density in
the Pomeron at low values of $\beta$ from HERA. Furthermore the Forward Proton 
Detector \cite{FPD}
installed by the D0 collaboration for Run II will be of great help to get a
direct measurement of the diffractive structure functions.

\ffig{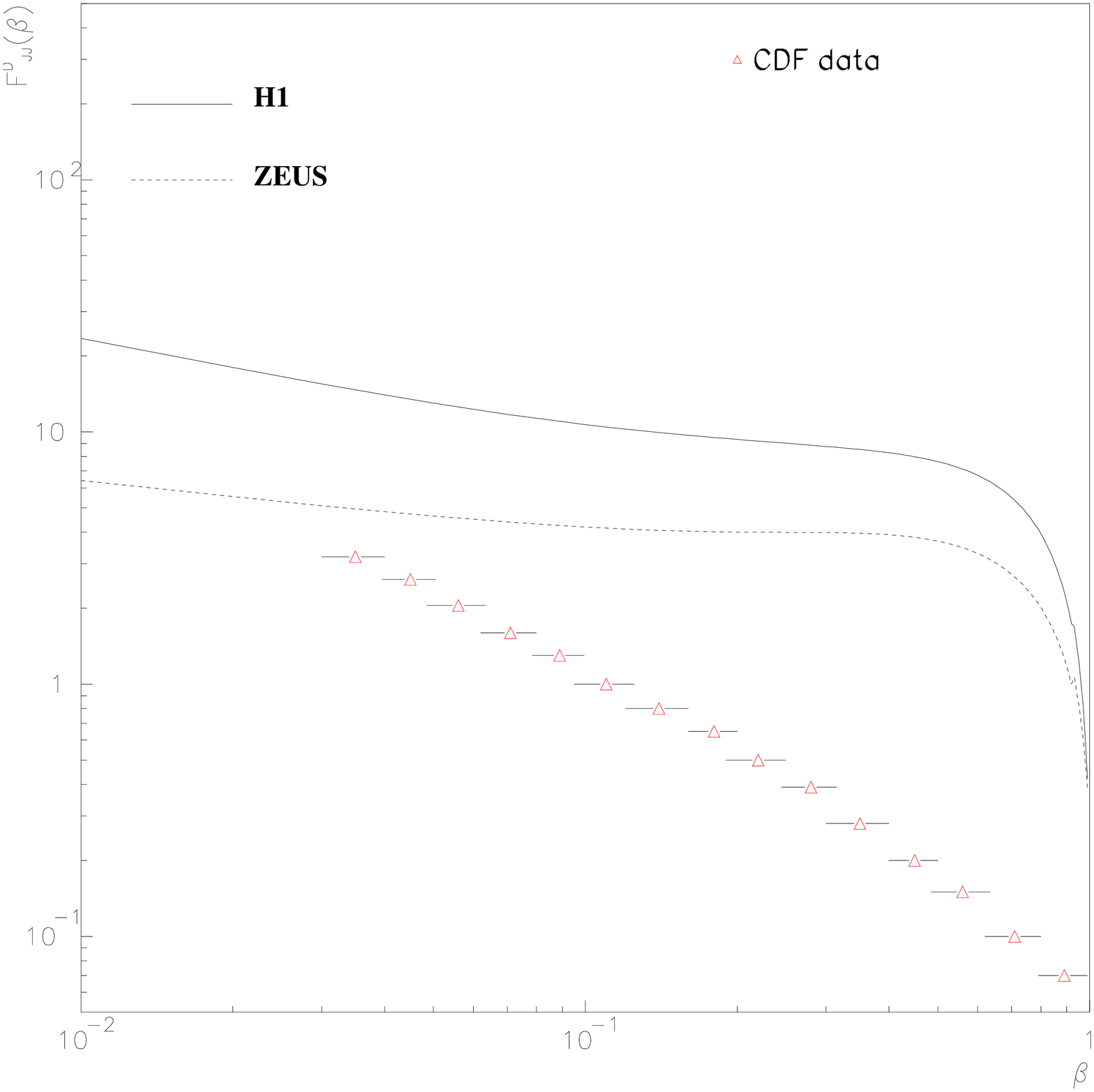}{130 mm}
{CDF data $\beta$ distributions compared with extrapolations from the
parton densities of the proton extracted from diffractive $F_2^{D(3)}$
measurements from H1 and ZEUS collaborations}
{cdf}

\section{Conclusion} 
Let us again go through the points discussed in the introduction: 

-We have developped a new  method to get the Pomeron and Reggeon 
exponents using Regge fits, confirming 
a higher value of the Pomeron intercept in ``hard'' diffraction than in soft processes
using a more precise method.

-We propose (and check the stability) of a new NLO DGLAP fit to the diffractive 
structure 
functions. We have obtained parton distributions for the Pomeron using H1
and ZEUS data. 

-H1 and ZEUS data both require a large gluon component inside the 
Pomeron but the gluon densities are quite different. The quark densities are 
found to be
similar in shape and magnitude. Using a parametrisation for the higher twist
contribution to $F_2^D$ allows us to get a description of the full set of data
using the NLO DGLAP evolution equations, and leads to the same parton 
distribution
as before. The longitudinal diffractive structure function measurement could
allow us to further constrain these higher twist parametrisations. 

-Differences
in the H1 and ZEUS gluon distributions  lead to significant differences in the 
longitudinal and charm contributions to $F_2^D$.
The low $Q^2$ extrapolation shows some scaling violations which do
not seem to be present in the data, which might be a sign of saturation effects.

-The same difference leads to diverging extrapolations for 
single diffraction at the Tevatron. While the H1-based extrapolation leads to a 
strong factorization breaking in both shape and normalization (at least a factor of 
10 with respect to CDF data), the ZEUS-based extrapolation leads to a possible 
compatibility with factorization at least in the small $\beta$ (large 
diffractive mass) range.

All these questions deserve a development of experimental and theoretical work 
in this domain, with the prospects of undestanding key aspects of QCD properties 
of ``hard'' diffraction. The possibility of diffractive production of heavy 
objects (jets, Higgses) at tevatron even enhances the interest of such studies.

\section{Acknowledgments}
We thank  A.Brandt, D.Goulianos, P.Marage, P.Newman, and P.Van Mechelen
for useful discussions and  a careful reading of the
manuscript.

\newpage

{\bf Appendix  A1: Decorrelation method to get the Pomeron and Reggeon 
intercepts}

\vspace{0.5cm}
Rewriting equation (\ref{reggeform}) as

\begin{eqnarray}
F_2^{D(3)}(Q^2_j,\beta_k,x_{\PO ,i}) =
f_{\PO / p} (x_{\PO ,i}) A_{jk}(Q^2_j,\beta_k)
+ f_{\RO / p} (x_{\PO , i}) B_{jk}(Q^2_j,\beta_k) 
\end{eqnarray}
A fit is then performed in each $(Q^2_j,\beta_k)$ bins.
$A$ and $B$ are only related to the $Q_j^2,\beta_k$ bin considered,
which means that they can be fitted within this bin
by minimizing the following $\chi^2_{jk}(Q^2_j,\beta_k)$ functions

We have 
$
\chi^2= \sum_j \sum_k \chi^2_{jk}(Q^2_j,\beta_k)
$ and

\begin{equation}
\chi^2_{jk}(Q^2_j,\beta_k)=
\sum_{i} \frac{(f_{\PO / p} (x_{\PO,i}) A_j(Q^2_j,\beta_k)
+ f_{\RO / p} (x_{\PO,i}) B_j(Q^2_j,\beta_k) -
F_2^{D(3),exper.}(Q^2_j,\beta_k,x_{\PO,i}))^2}
{\Delta F_2^{D(3),exper.}(Q^2_j,\beta_k,x_{\PO,i}))^2}
\end{equation}
where the $i$ index runs over the number of points
with different values of $x_{\PO}$ in the $Q^2_j,\beta_k$ bin considered
and $\Delta F_2^{D(3),exper.}$ is the statistical error on $F_2^D$ (denoted
$\sigma_i$ in the following).
We use the notations 

\begin{eqnarray}
&~& f_{\PO / p} (x_{\PO,i})=f_{\PO,i} \\
&~& f_{\RO / p} (x_{\RO,i})=f_{\RO,i} \\
&~& F_2^{D(3),exper.}(Q^2_j,\beta_k,x_{\PO,i})=F_i.
\end{eqnarray}

As $A_{jk}$ and $B_{jk}$
do not depend on $x_{\PO,i}$, they can be calculated directly
by solving the following system

\begin{eqnarray}
\frac{\partial \chi^2_j(Q^2_j,\beta_k)}{\partial A_{jk}} &=& 0 \\
\frac{\partial \chi^2_j(Q^2_j,\beta_k)}{\partial B_{jk}} &=& 0 
\end{eqnarray}
which gives

\begin{eqnarray}
A_{jk} &=& \frac{
\sum_i \frac{2 f_{\PO,i} F_i}{\sigma^2_i} 
\sum_i \frac{2 f_{\RO,i}^2 }{\sigma^2_i}
-
\sum_i \frac{2 f_{\RO,i} F_i}{\sigma^2_i} 
\sum_i \frac{2 f_{\RO,i} f_{\PO,i} }{\sigma^2_i} 
}
{
\sum_i \frac{2 f_{\PO,i}^2 }{\sigma^2_i} 
\sum_i \frac{2 f_{\RO,i}^2 }{\sigma^2_i}
-
(\sum_i \frac{2 f_{\RO,i} f_{\PO,i}}{\sigma^2_i})^2
}
\\
B_{jk} &=& \frac{
\sum_i \frac{2 f_{\RO,i} F_i}{\sigma^2_i} 
\sum_i \frac{2 f_{\PO,i}^2 }{\sigma^2_i}
-
\sum_i \frac{2 f_{\PO,i} F_i}{\sigma^2_i} 
\sum_i \frac{2 f_{\RO,i} f_{\PO,i} }{\sigma^2_i} 
}
{
\sum_i \frac{2 f_{\PO,i}^2 }{\sigma^2_i} 
\sum_i \frac{2 f_{\RO,i}^2 }{\sigma^2_i}
-
(\sum_i \frac{2 f_{\RO,i} f_{\PO,i}}{\sigma^2_i})^2
}
\end{eqnarray}

Thus, for each $Q^2_j,\beta_k$ bin, $A_{jk}$ and $B_{jk}$ are well defined 
values
and depend on $\alpha_{\PO}(0)$ and $\alpha_{\RO}(0)$
via the flux factors.

In conclusion, all $A_{jk}$ and $B_{jk}$ parameters should not be introduced 
without
care as free parameters
in the global Regge fit as they satisfy the two last expressions in terms
of $F_i$, $f_{\PO,i}$ and $f_{\RO,i}$, which reduces considerably
the number of free parameters. Indeed only $2$ free parameters remain~:
$\alpha_{\PO}(0)$ and $\alpha_{\RO}(0)$.

\newpage

{\bf Appendix A2: ZEUS and H1 fit parameters}

\vspace{0.5cm}
$F_2^{D(3)}$ can be expressed as a sum of two factorized 
contributions corresponding to a Pomeron and secondary Reggeon trajectories.
The Pomeron and Reggeon fluxes are assumed to follow a Regge behaviour with  
linear
trajectories $\alpha_{\PO,\RO}(t)=\alpha_{\PO,\RO}(0)+\alpha^{'}_{\PO,\RO} t$, 
such
that

\begin{equation}
f_{{\PO} / p,{\RO} / p} (x_{\PO})= \int^{t_{min}}_{t_{cut}} 
\frac{e^{B_{{\PO},{\RO}}t}}
{x_{\PO}^{2 \alpha_{{\PO},{\RO}}(t) -1}} {\rm d} t 
\nonumber
\end{equation}
where $|t_{min}|$ is the minimum kinematically allowed value of $|t|$ and
$t_{cut}=-1$ GeV$^2$ is the limit of the measurement. 

\begin{eqnarray}
F_2^{D(3)}(Q^2,\beta,x_{\PO})=
f_{\PO / p} (x_{\PO}) F_2^{\PO} (Q^2,\beta)
+ N_{\RO} f_{\RO / p} (x_{\PO}) F_2^{\RO} (Q^2,\beta) \ .
\nonumber
\end{eqnarray}

The values of
$\alpha_{{\PO},{\RO}}(0)$ are free parameters while
$B_{{\PO},{\RO}}$ and $\alpha^{'}_{{\PO},{\RO}}$ are taken from hadron-hadron 
data ($\alpha^{'}_{\PO}=0.26$ GeV$^{-2}$, $\alpha^{'}_{\RO}=0.90$ GeV$^{-2}$,
$B_{\PO}=4.6$ GeV$^{-2}$, $B_{\PO}=2.0$ GeV$^{-2}$
The Pomeron intercept coming from Regge fits is $\alpha_{\PO}(0)=1.20 \pm 0.02$
for H1, and $\alpha_{\PO}(0)=1.13 \pm 0.04$ for ZEUS.

The Reggeon structure function is needed only to describe H1 data, and
the value of $\alpha_{\RO}(0)=0.62 \pm 0.02$,
with the normalization $N_{\RO}=14.2 \pm 0.1$. The values of
$F_2^{\RO} (Q^2,\beta)$ are taken from a parameterisation of the 
pion structure function \cite{GRVpion}.

The parametrisations for the gluon and quark densities 
($z{ {S}}_{q}(z,Q^2)=u+\bar{u}+d+\bar{d}+s+\bar{s}$) in the Pomeron
are given at $Q_0^2$=3 GeV$^2$:

\begin{eqnarray}
z{\it {S}}(z,Q^2=Q_0^2) &=& \left[
\sum_{j=1}^n C_j^{(S)} \cdot P_j(2z-1) \right]^2
\cdot e^{\frac{a}{z-1}} \\
z{\it {G}}(z,Q^2=Q_0^2) &=& \left[
\sum_{j=1}^n C_j^{(G)} \cdot P_j(2z-1) \right]^2
\cdot e^{\frac{a}{z-1}}
\label{gluon1}
\end{eqnarray}
where 
$P_j(\zeta)$ is the
$j^{th}$ member in a set of Chebyshev polynomials, which are chosen such that
$P_1=1$, $P_2=\zeta$ and $P_{j+1}(\zeta)=2\zeta P_{j}(\zeta)-P_{j-1}
(\zeta)$.

\begin{center}
\begin{tabular}{|c|c|c|} \hline
 parameters & H1 &  ZEUS  \\ 
\hline\hline
 $C_1^{(S)}$ & 0.18     $\pm$ 0.05 &  0.41  $\pm$0.02 \\
 $C_2^{(S)}$ & 0.07 $\pm$ 0.02 & -0.16  $\pm$ 0.03 \\
 $C_3^{(S)}$ & -0.13     $\pm$ 0.02 & -0.11  $\pm$ 0.02 \\ \hline
 $C_1^{(G)}$ & 0.82      $\pm$ 0.40 & 0.53     $\pm$ 0.30 \\
 $C_2^{(G)}$ & 0.22 $\pm$ 0.06 & 0.28 $\pm$ 0.25 \\
 $C_3^{(G)}$ & 0.01 $\pm$ 0.04 & 0.02 $\pm$ 0.11
 \\
\hline
\end{tabular}
\end{center}
\vskip .5cm
\begin{center}
{Table I- Pomeron quark and gluon densities parameters}
\end{center}

The parametrisation for the  higher-twist contribution to the diffractive 
structure function is the following

\begin{eqnarray}
 F_2^{D(3),III} = C \left( \frac{x_0}{x_{\PO}}
\right)^{n_4}  \left( \frac{Q_0^2}{Q^2} \right)
\left( ln (\frac{Q^2}{4 Q_0^2 \beta} +1.75) \right)^2
\beta^3 (1-2 \beta)^{2}
\end{eqnarray}
where
\begin{eqnarray}
n_{2,4}= n^0_{4} + n^1_{4} ln \left[ ln \left( \frac{Q^2}{Q_0^2} \right)
+1 \right]
\end{eqnarray}

with the following values for the parameters

\begin{center}
\begin{tabular}{|c|c|c|} \hline
 parameters & H1 &  ZEUS  \\ 
\hline\hline
$C$ & 0.035 $\pm$ 0.016 &  357.  $\pm$ 106. \\
$x_0$ & 0.40 $\pm$ 0.02 & 0.0008 $\pm$ 0.0002 \\
$n_4^0$ & 1.43 $\pm$ 0.08  & 1.00 $\pm$ 0.10 \\
$n_4^1$ & 0.00 $\pm$ 0.05  &  0.37  $\pm$ 0.10 
 \\
\hline
\end{tabular}
\end{center}
\vskip .5cm
\begin{center}
{Table II: Parameters obtained for the higher-twist parametrisation}
\end{center}

\newpage


\end{document}